\documentstyle[prd,aps]{revtex}
\begin{document}

\title{Unstable states in QED of strong magnetic fields\footnote{%
hep-ph/9601229. To appear in {\em Physical Review D53~(2), 1996.}}}

\author{M. Kachelrie{\ss}\footnote{email: Kachelriess@lngs.infn.it}}

\address{INFN, Laboratori Nazionali del Gran Sasso, 
         I--67010 Assergi (AQ), Italy}

\address{Theoretische Physik I, Ruhr-Universit\"at Bochum, 
         D--44780  Bochum, Germany}

\maketitle

\begin{abstract}
 We question the use of stable asymptotic scattering states in QED 
 of strong magnetic fields. To correctly describe excited Landau 
 states and photons above the pair creation threshold the asymptotic 
 fields are chosen as generalized Licht fields. In this way the off-shell 
 behavior of unstable particles is automatically taken into account, 
 and the resonant divergences 
 that occur in scattering cross sections in the presence of a strong 
 external magnetic field are avoided. 
 While in a limiting case the conventional electron propagator with
 Breit-Wigner form is obtained, in this formalism it is 
 also possible to calculate
 $S$-matrix elements with external unstable particles. 
\end{abstract}


\section{Introduction}

The discovery of neutron stars with extremely strong magnetic fields 
$B$ up to
$10^{13}$ Gauss has given the impetus to numerous calculations of QED 
processes in which the magnetic field is taken into account exactly. 
A remarkable feature of these magnetic field strengths is that the 
cyclotron energy becomes of the order of the electron rest energy, 
and, consequently, the quantization of the electron states into 
{\em discrete\/} Landau levels
becomes important. These QED processes were 
recalculated using conventional perturbation theory in the Furry picture:
the free electron propagator and the free wave function of an
electron were replaced by the exact Green's function and the exact 
solution of the Dirac equation with a homogenous, static magnetic field. 
Then the same Feynman rules were applied. In this way, most first- and 
second-order processes have been recalculated in the last fifteen
years (for a review cf. Ref. \cite{ha91}).
However, the results often show an unsatisfactory behavior.
Because of strict energy conservation, first-order processes such as
cyclotron absorption have delta-function-like decay-widths 
or, like cyclotron emission, become infinite averaging
over ``reasonable''
distribution functions. This is remedied by accounting for the finite
lifetime of the {\em external\/} states; i.e., 
one replaces the delta-function 
expressing energy conservation by a Lorentz curve in the decay-width. 
In the case of second- or higher-order processes 
singularities arise due to
on-shell intermediate states. Here, one accounts for the finite
lifetime of the {\em intermediate\/} Landau states and replaces in the
electron propagator Feynman's
$i\varepsilon$ with $\frac{1}{2}i\Gamma_{N,\tau}$, where 
$\Gamma_{N,\tau}$ is the decay-width
of the electron state with Landau quantum number $N$ and
polarization $\tau$. 

Despite the use of the electron propagator with a complex mass
some processes still lead to divergent cross sections. The most
prominent example is magnetic Compton scattering with an initial photon
which is above the pair creation threshold. If the intermediate
electron is in the stable Landau ground state, there is no decay-width
associated with it,
and, consequently, the total cross section is divergent everywhere
above the pair creation threshold \cite{he79}.
Another problem arises if more than one particle is unstable.
Then it is not obvious how the decay-widths should be defined. 
Usually, for first-order processes the total decay-width
is assumed to be additive, i.e., to be the sum of the decay-widths
of the individual particles. In contrast, for second-order
processes the decay-width of 
every single virtual particle  is chosen as its individual
on-shell decay-width. This seems to be arbitrary and 
shows the absence of a comprehensive strategy to treat the
instability of electrons and photons in magnetic fields.
Therefore, it is the purpose of this paper to formulate a well-defined
perturbation theory for QED of strong magnetic fields 
where the finite lifetime of excited Landau states and photons above
the pair creation threshold is automatically incorporated. 
To this end, we give up the concept of stable scattering states and
instead introduce generalized Licht fields for the unstable particles.
The energy of particles described by Licht fields is not fixed by an
on-shell condition but is given by some spectral function. We do not
attempt to calculate these spectral functions from first principles
because they are well approximated for practical calculations 
by Lorentz curves \cite{ja95}. The advantage of using Licht fields
is that within this formalism it is possible to take into account
consistently the instability of intermediate {\em and} external 
particles.

As an application we show, for the generic example of magnetic
Compton scattering in which the initial photon is above the
pair creation threshold, how the use of Licht fields eliminates
resonant divergences of QED with strong magnetic fields.

%%%%%%%%%%%%%%%%%%%%%%%%%%%%%%%%%%%%%%%%%%%%%%%%%%%%%%%%%%%%%%%%%
\section{Unstable states and propagators}

To see the underlying reasons for the unsatisfactory behavior of cross
sections of QED with strong magnetic fields, we remind the reader of two 
failures of perturbation theory in the Furry picture for $B>0$
using stable particle states and propagators:

i)  
 It is a highly distinctive feature of vacuum theory that the Hilbert
 space structure does not change in going from the free to the 
 interacting theory. This remains true for $B>0$ only if the electron
 self-energy and the vacuum polarization
 vanish on-shell. But this is not the case due to the 
 imaginary part of the self-energy indicating the decay of 
 Landau states for $N>0$ 
 and of the vacuum polarization indicating
 the decay of a photon in an $e^- e^+$-pair above the pair creation
 threshold \cite{selbst,mass}. 

% ii) 
% It is not possible to implement the unitary time evolution 
% $\Psi(t')=U(t' ,t)\Psi(t)$ for the electron field $\Psi$ 
% in Fock space in the case of external magnetic fields,
% in contrast to external electric fields \cite{ru}. 
% One way to interpret this result is to seek the analogy to 
% finite temperature quantum field theory: there, the 
% Narnhofer-Requardt-Thirring
% theorem states the impossibility of finite temperature perturbation
% theory in terms of stable fields \cite{nt} and a consistent theory
% must use unstable states \cite{la}. 

ii) 
 Fields that describe unstable states have a vanishing LSZ-limit 
 \cite{li,lu}.
 
%These rigorous results are generally ignored by the community applying 
%this theory. 

For us these results are the starting point for identifying the 
correct states and propagators for decaying electrons and photons in 
strong magnetic fields. There are two ways of introducing unstable 
particles in quantum field theory.
Usually, the notation of a complex mass shell, $p_\mu p^\mu= m^2-i\Gamma$,
is used. This %ansatz 
approach is easily applied for propagators yielding 
the typical Breit-Wigner shape for resonances in cross sections \cite{lo}, 
but was recently 
also generalized to external lines \cite{gr95}.
Another method is the use of generalized Licht fields. Here, one
replaces the on-shell condition $p_\mu p^\mu= m^2$ by an off-shell
mass spectral density. The abandoning of the on-shell condition
is justified by the time-energy uncertainty which forbids an unstable 
particle to have a fixed energy. 
The use of Licht fields is additionally motivated by the following
two reasons:
First, generalized Licht fields have non-vanishing
LSZ-limits. Second, in the case of propagators, the Licht field
approach is the more general one and contains 
the complex mass shell method as a special case. Therefore,
we follow the second approach in this work and,
in the spirit of Refs. \cite{li,lu,la}, introduce suitable 
generalized Licht fields for the unstable particles. 
However, the two different methods
result in a different treatment of external lines. These differences
will be discussed at the end of Section \ref{spe}.

For the electron we define the Licht field by
\begin{equation}       \label{def}
 \Psi (x) =  \sum _{a} \int_{0}^\infty dE 
  \left(  Z_{+,n}^{1/2} (E) b_{a} (E) \psi^{(+)}_{a}({\bf x})e^{-iEt} +
    Z_{-,n}^{1/2} (E)d^{\dagger}_{a} (E) \psi^{(-)}_{a}({\bf x})e^{+iEt} 
  \right)  
\end{equation}
%
%\begin{equation}       \label{def}
% \Psi (x) =  \sum _{a} \int_{0}^\infty dE \: Z_n^{1/2} (E)
 % \left( b_{a} (E) \psi^{(+)}_{a}({\bf x})e^{-iEt} +
 %d^{\dagger}_{a} (E) \psi^{(-)}_{a}({\bf x})e^{+iEt} \right)  \; , 
%\end{equation}
%
and $\bar\Psi=\Psi^\dagger\gamma^0$,
where the anticommutation relations are the usual ones for $N=0$; 
for $N>0$
\begin{equation}
 \left\{ b_{a} (E) , b_{a'}^\dagger (E') \right\} = 
 \left\{ d_{a} (E) , d_{a'}^\dagger (E') \right\} = 
 \delta_{a,a'} \delta (E-E')
\end{equation}
and the other anticommutators are zero. Here, $\psi^{(\lambda)}_{a}$ 
are the energy solutions of the Dirac equation
in the presence of the external magnetic field ${\bf B}=B {\bf e}_z$,
$a=\{N,\tau,p_y,p_z\}$ denotes the set of
quantum numbers needed in order to completely characterize the
solutions, and $\lambda =\pm$ distinguishes positive and negative 
energy solutions \cite{so,cyc}.
The energy $E$ of the particle is smeared around the on-shell value
$E_n =\sqrt{m^2 +2NeB +p_z^2}$ due to the integration over the
spectral functions $Z^{1/2}_{\lambda, n} (E)$. 
These functions are generalizations of the wave function 
renormalization constant $Z$ of a stable field and labeled
by that subset of quantum numbers $n=\{N,\tau\}$ which enters in the 
decay-width of the unstable states. In the following, we will suppress
the other quantum numbers $p_y$ and $p_z$ \cite{p_z}. The dependence of
$Z_{\lambda,n}^{1/2}$ on $\lambda$ reflects 
the different time evolution of positive and negative energy solutions. 
Therefore, the difference between $Z_{+,n}^{1/2}$
and $Z^{1/2}_{-,n}$ should show up only as some kind of 
boundary condition.  
In order to get a charge symmetric theory, the condition 
$Z^{1/2 \ast}_{+,n}= e^{i\phi} Z^{1/2}_{-,n}$, where $\phi$ is
an arbitrary phase, follows.

A physical interpretation of Eq. (\ref{def}) is that $\Psi$ describes
$n$-times different particles, i.e., every Landau state with distinct
$N$ and $\tau$ would be identified as a different particle. 
%only the ground state $n=(N=0,\tau = -1)$
%is stable. It has $Z_{0,-1}^{1/2}=\delta (E-E_0 )$ and the usual 
%LSZ-limit $\psi_{0,-1}\to Z_2 \psi_{0,-1}^{in}$
%for $t\to -\infty$. Here, $Z_2$ is the normal electron wave function
%renormalization {\em constant\/} and the limit -- as all of the 
%following -- should be understood in the weak operator topology. 
%For $N>0$, the particles can decay and $Z_n(E)$ weights the 
%contributions 
%of creating an unstable ``$(N,\tau)$-particle'' with energy $E$. 
%
Excited states with $N>0$ are unstable
because of the interaction with the photon field.
These particles can decay and $Z_{\lambda,n}^{1/2}(E)$ 
weights the contributions 
of creating an unstable ``$(N,\tau)$-particle'' with energy $E$ and 
polarization $\tau$.
Only the ground state $n=(N=0,\tau = -1)$ remains stable. To recover
the usual on-shell energy relation and wave function of a 
stable electron, one has to set $Z_{\lambda,0,-1}^{1/2}=\delta (E-E_0 )$.
Then, the ground state has
the usual LSZ-limit $\psi_{0,-1}\to Z_2^{1/2} \psi_{0,-1}^{{\rm out}}$
for $t\to\infty$, where $Z_2$ is the normal electron wave function
renormalization {\em constant\/} and the limit -- as in all of the 
following -- should be understood in the weak operator topology.

In the case of the photon field with its continuous energy spectrum
we adopt, in an analogous way, the Licht field
\begin{equation}
 A_\mu (x)= \int_{0}^\infty ds' \left(
            Z_{+,r}^{1/2} (s') A^{(+)}_\mu (x,s') 
          + Z_{-,r}^{1/2} (s') A^{(-)}_\mu (x,s') \right) \; ,
\end{equation}
where $A^{(+)}$ and $A^{(-)}$ are the positive and negative energy
photon fields, respectively.
Similarly to the case of the electron field, the
functions $Z_{\lambda,r}^{1/2} (s')$ are labeled besides by $\lambda$
by those quantum numbers on which the 1-$\gamma$-pair production
probability $\Gamma_r (s)$ depends: the energy perpendicular to the
magnetic field $s=\omega\sin\theta$, and the polarization $r$ of
the photon \cite{pair}. The part of the
photon field $A^{(\lambda)}_\mu (x)$ with energy
below the pair creation threshold $\omega=2m/\sin\theta$ 
is stable. Therefore, for $s<2m$, the functions 
$Z_{\lambda, r}^{1/2} (s')=\delta(s-s')$ and the field 
has the usual LSZ-limit
$A^{(\lambda)}_\mu (x,s) \to Z_3^{1/2} A^{{\rm out}(\lambda)}_\mu (x,s)$ 
for $t\to\infty$.

In contrast with the fields $\Psi$ and $A^\mu$ describing unstable 
particles, the component fields $A^{(\lambda)} (x,s')$ and
\begin{eqnarray}
 \psi^{(+)}_{n}(x,E) & = & 
  b_{n}(E) \psi^{(+)}_{n}({\bf x})e^{-iEt} 
\\
 \psi^{(-)}_{n}(x,E) & = & 
  d_{n}^{\dagger} (E) \psi^{(-)}_{n}({\bf x})e^{+iEt}
\end{eqnarray}
do have a non-vanishing LSZ-limit \cite{li,lu}:
\begin{equation}
 A^{(\lambda)}_\mu (x,s') \to   A^{(\lambda){\rm out}}_\mu (x,s')
 \qquad\mbox{for}\quad
 t \to \infty 
\end{equation}
and
\begin{equation}        \label{LSZ}
 \psi^{(\lambda)}_{n} (x,E) \to  \psi^{(\lambda)\rm out}_n (x,E) 
%  b_{n}^{\rm in} (E) \psi^{(+)}_{n}({\bf x})e^{-iEt} +
% d_{n}^{\dagger , {\rm in}} (E) \psi^{(-)}_{n}({\bf x})e^{+iEt}
 \qquad\mbox{for}\quad
 t \to \infty \; .
\end{equation}
Therefore, 
one is able to compute Green's functions with the Gell-Mann--Low 
or the LSZ-reduction formula using the decomposed out-fields. 
Expressed in terms of these, the interaction Hamiltonian $H_I$ reads
\begin{eqnarray}
 H_I (t) & = &  
 -e \int d^3 x \sum_{n_1 ,n_2} \sum_{\lambda_1,\lambda_2, \lambda'}
 \int_{0}^{\infty} dE_1 
 \int_{0}^{\infty} dE_2  \: \int_{0}^{\infty} ds' 
 Z_{\lambda_1,n_1}^{1/2} (E_1) \: Z_{\lambda_2,n_2}^{1/2} (E_2) 
 \: Z_{\lambda' ,r}^{1/2} (s') 
\nonumber\\ & &
 \overline{\psi}_{n_1}^{(\lambda_1) \rm out} (x, E_1) \gamma^\mu 
 \psi_{n_2}^{(\lambda_2) \rm out} (x, E_2) 
 A_{\mu}^{(\lambda') \rm out} (x,s')  \; ,
\end{eqnarray}
where we omit all counterterms. Since here we are only interested in
first and second-order processes, we do not take care of 
renormalization. But we want to mention that, since physical
parameters should be chosen as directly observable quantities,
the bare parameter $m_0$ should be expressed by the physical mass
$m$ only for the Landau ground state. In this way, the usual electron
wave function renormalization constant $Z_2$ is fixed. However, for
decaying states one should choose as a physical parameter instead of
$m$  some characteristic parameter of an unstable state, e.g. the
decay-width $\Gamma_n$. 

Now, we are ready to derive diagrammatic perturbation theory.
As a first step we compute the components of the electron propagator
(omitting the index ``out'' from now on) for $N>0$, 
\begin{eqnarray}
 \lefteqn{ iS_F (x_1,E_1,n_1 ;x_2,E_2,n_2 ) =
 <0| T \left( \psi_{n_1}^{(\lambda)}(x_1,E_1) 
        \overline{\psi}_{n_2}^{(\lambda)}(x_2,E_2)\right)|0 >  
  = } \nonumber\\ &  & =
 \delta (E_1 -E_2) \delta_{n_1,n_2}
 \int_{-\infty}^{\infty} \frac{ds}{2\pi} 
 \frac{1}{s- \lambda\left( E_1 -i\varepsilon \right)} \:
 \psi^{(\lambda)}_{n_1}({\bf x_1})
 \overline{\psi}^{(\lambda)}_{n_2}({\bf x_2}) \: 
 e^{-is(t_1-t_2)}  \; , 
\end{eqnarray}
and read off the vertex as 
\begin{equation}
 ie\gamma^\mu  Z^{1/2}_{\lambda_1 ,n_1} (E_1) 
               Z^{1/2}_{\lambda_2 ,n_2} (E_2) 
               Z^{1/2}_{\lambda' ,r} (s') \; .
\end{equation}
Using the decomposed propagator, one has not only to integrate over all        
not fixed momenta, but also over the variables of the functions $Z$.
However, we are mainly interested in Green's functions of $\Psi$. 
The total propagator is sandwiched between two vertices. One of the two
integrations over $E$ breaks down due to the delta-function and
one obtains 
\begin{eqnarray}
 \lefteqn{ iS_F (x_1 ,x_2 ) =
 <0| T \left( \Psi(x_1) \overline{\Psi}(x_2)\right)|0 >  
  = } \nonumber\\ &  & =
 \sum _{n,\lambda}  \int_0^\infty dE \: Z_n (E) 
 \int_{-\infty}^{\infty} \frac{ds}{2\pi}
 \frac{1}{s- \lambda\left( E -i\varepsilon \right)} \:
 \psi^{(\lambda)}_{n}({\bf x_1})
 \overline{\psi}^{(\lambda)}_{n}({\bf x_2}) \: e^{-is(t_1-t_2)}  \; . 
\end{eqnarray}
Here, we set the ill-defined $Z_{0,-1} (E)$ equal to 
$\delta (E-E_{0})$ to obtain a compact expression 
and the vertex becomes the usual $ie\gamma^\mu$. 
The functions $Z_n(E)$ are abbreviations for $|Z_{\lambda,n}^{1/2}|^2$.
Therefore, they are real and independent from $\lambda$.
One should remember that the propagator obtained, although similar
to the spectral representation of the full propagator in vacuum 
theory, is a bare one.

The derivation of the photon propagator is similar.
However, there arises the additional difficulty that the usual
spin projection operators do not work for off-shell states:
a propagator for spin-$s$-particles
will generally contain particles with lower spin values 
$(s-1,s-2,\ldots ,0)$. But since Feynman diagrams with virtual photons 
do not play a prominent role in the astrophysical applications, we omit 
the derivation here \cite{spin}.

\section{Spectral functions}   \label{spe}

In order to make the whole treatment consistent,
the spectral functions $Z_n (E)$ and $Z_r (s)$
have to be -- at least in principle --
computable. A hard way is to use the fact that the 
$Z^{1/2}_{\lambda,n} (E)$ completely
determine through the Eqs. (\ref{def}) and (\ref{LSZ})
the normalization of the components of the field $\Psi$. The latter
is fixed by the canonical anticommutation relations. Therefore
one can use perturbation theory in the Heisenberg picture to
calculate $Z^{ 1/2}_{\lambda,n} (E)$, and, similarly, 
$Z^{ 1/2}_{\lambda,r} (s)$  \cite{lu,la}. 

In practice, these calculations are nearly intractable and one will 
use a guess. 
The {\it Ans{\"a}tze} according to conventional wisdom are Lorentzians,
\begin{equation}
 Z_n^L (E) = 
 \frac{\Gamma_n}{\pi\left( \left( E-E_n \right)^2 + 
      \frac{1}{4}\Gamma_n^2 \right)}  
\end{equation}
for the electron and
\begin{equation}
 Z_r^L (s') = 
 \frac{\Gamma_r (s)}{\pi\left( \left(s-s' \right)^2 + 
      \frac{1}{4}\Gamma_r^2 (s') \right)}  
\end{equation}
for the photon, where we choose $\Gamma_n$ and $\Gamma_r(s)$
to be the total decay-width of the Landau level
$(N,\tau)$ \cite{cyc} and the 1-$\gamma$-pair production 
probability $\Gamma_r (s)$ \cite{pair},
respectively, calculated in conventional perturbation theory.

This {\it Ansatz} reproduces the electron propagator with the normally used 
Breit-Wigner 
prescription $i\varepsilon\to\frac{1}{2}i\Gamma_n$ for $\Gamma_n\ll E_n$.
In this case, after expanding the poles, the lower limit of
integration can be extended from $0$ to $-\infty$
producing two poles at $s-\lambda (E_n-\frac{1}{2}i\Gamma_n)$.
According to Ref. \cite{gr93},  the Breit-Wigner approximation
yields a result that is always consistent within the perturbation theoretical
order of calculation. In particular, the author showed that for all $B$ 
and $N$ the deviations from the Breit-Wigner line shape are small.
From our derivation follows the usual restriction $\Gamma_n \ll E_n$  
for the validity of the approximation.
%This corresponds to the usual condition $\Gamma^2 \ll q^2$
%in field-free quantum field theory, where $q$ is the propagator momentum. 
Using the approximative formulae of Ref. \cite{pa}
for $\Gamma_n$, one sees that the condition $\Gamma_n \ll E_n$
is indeed always satisfied.

However, from a more fundamental point of view, it is clear that
the extension of the integration to negative $E$ violates the 
spectral condition and is at the root of the violation of unitarity 
and causality.
Although therefore the consistency of the approach is lost, it seems
to us worthwhile to explore the consequences of this {\it Ansatz}.
In contrast with the usual derivation
of the propagator with Breit-Wigner shape,
the derivation presented here gives a complete
scheme that describes unstable particles 
as external as well as intermediate states.  
Furthermore, one can treat 
processes with more than one unstable particle 
without ambiguities. Therefore,  in this section
we do not attempt to calculate
the spectral functions $Z$ from first principles but restrict
ourselves to the simpler task of investigating the consequences
of the Breit-Wigner approximation %approximation $Z=Z_L$ 
in the Licht field approach.

Denoting the electron wave functions obtained in this approximation 
by $\psi^{(\lambda)}_{L,n} (x)$, we obtain
\begin{equation}   
 \psi^{(\lambda)}_{L,n} (x) = N \,\psi^{(\lambda)}_{n}({\bf x}) \,
                              e^{-i\lambda E_n t - \frac{1}{2}\Gamma_n t} 
                              \theta (t)
\end{equation}
and $\bar\psi^{(\lambda)}_{L,n}=\psi^{(\lambda)\dagger}_{L,n}\gamma^0$.
Here, $N$ is a normalization constant and we chose the signs in
\begin{equation}
 Z_{\lambda,n}^{1/2} = \sqrt{\frac{\Gamma_n}{\pi}} 
               \frac{i\lambda}{E-E_n + \frac{1}{2}i\lambda\Gamma_n}
\end{equation}
according to the following two requirements: first, we demand that the
wave functions do not vanish for $t>0$ and, second, 
% they should be Dirac conjugate to each other. Third, 
in the limit $\Gamma_n\to 0$ the phase of $\psi_{L,n}^{(\lambda)}$ has to
coincide with the phase of $\psi_{n}^{(\lambda)}$.
The first requirement leads automatically to decaying 
states, for both positive
and negative energy solutions. The choice of 
non-vanishing wave functions for $t<0$ results in states 
whose norm grows in time. Therefore, the choice between
non-vanishing wave functions for $t>0$ or $t<0$ corresponds to
the choice of the direction of the time arrow and has to be made
by hand. As anticipated, the different form of $Z_{\lambda,n}^{1/2}$
for $\lambda = \pm$ is necessary to obtain the correct boundary
condition for decaying states. 
Similarly, we obtain for the photon wave functions
\begin{equation}
 A^{(+,r)}_{L,\mu} (x) =  N \left( 2\omega V \right)^{-1/2}
 \varepsilon^{(r)}_\mu \,
 e^{-i (\omega t - {\bf kx})} e^{- \frac{1}{2} \Gamma_r (\omega)t}
 \theta (t)   
\end{equation}
and $A^{(-,r)}_{L,\mu} = A^{(+,r)\ast}_{L,\mu}$.

Since the norm of the states varies with time, the correct normalization
is not obvious. A reasonable prescription is the requirement 
that cross sections calculated with decaying states coincide in the limit
$\Gamma\to 0$ with the same cross sections calculated in the usual
formalism.

Formally, $S$-matrix elements with external unstable particles 
will be calculated in the usual way. But 
because of the $\theta$ functions, the time integration over vertices with 
external unstable particle goes effectively only from $0$ to $\infty$.
Thereby, no divergent time integrals will be caused by the real part
of the exponentials of decaying states.

Finally, we want to compare our approach with that of a recent paper
\cite{gr95}. The authors propose the use of propagators with complexified
energies and -- this is the main difference -- the use of external states
\begin{eqnarray}
 \psi_{L,n}^{(\lambda)} (x) & = & 
         \psi_n^{(\lambda)} ({\bf x}) \, e^{-i\lambda E_n t 
            - \frac{1}{2}\lambda\Gamma_n t} 
\\
 \bar\psi_{L,n}^{(\lambda)} (x) & = & 
         \bar\psi_n^{(\lambda)} ({\bf x}) \, e^{+i\lambda E_n t
             + \frac{1}{2}\lambda\Gamma_n t} 
\end{eqnarray}
for the electrons and similar ones for photons. In contrast to our wave 
functions, the norm of the negative energy solutions grows in time.
Furthermore, these wave functions are valid for $t<0$ as well as
for $t>0$. Therefore, the real part of the 
exponentials will lead to divergent time integrals. Consequently,
no scattering amplitudes in the normal sense (i.e. 
for transitions between $t_{i,f}\to\pm\infty$)
with external unstable states can be calculated in this approach. 
Instead, one calculates matrix elements of the time evolution
operator $U(t_f,t_i)$ which are dependent on
the time lapse $t_f-t_i$ 
between preparation of the initial state and measurement
of the final state. The authors of Ref. \cite{gr95} claim that
this approach, which is the direct transcription of the
Wigner-Weisskopf method of non-relativistic quantum mechanics, eliminates  
all divergences of QED with strong magnetic fields. 
However, the main object of field theory, the $S$-matrix, is not 
generally computable in this formalism. Moreover, the wave function
$\bar\psi$ is not 
the Dirac conjugate spinor of $\psi$ since the real part of the exponent 
changes sign. Consequently, charge symmetry is lost and the $S$-matrix 
elements lack crossing symmetry. 
Finally, a more practical objection seems to be important. 
Since the main application of
QED of strong magnetic field is astrophysics, the usefulness
of cross sections which are dependent on
the time lapse between ``preparation'' and ``measurement''
of the states is restricted.

%%%%%%%%%%%%%%%%%%%%%%%%%%%%%%%%%%%%%%%%%%%%%%%%%%%%%%%%%%%%%%%%%%%%%%%%
\section{Applications}          \label{app}

In this section,
we want to illustrate some basic consequences of this formalism.
 
First, we consider a generic first-order process. 
The $S$-matrix element is given by
\begin{equation}
 S_{fi}=
 \int dE_1 dE_2 ds' \: Z_{1}^{1/2} (E_1) 
 Z_{2}^{1/2} (E_2) Z^{1/2}_r (s')  S_{fi}^{\rm conv}  \; ,    
\end{equation}
where $S^{\rm conv}_{fi}$ is the conventional $S$-matrix element
but with off-shell energies. 
Choosing the spectral functions to be Lorentz curves,
there is no difference between the approach presented in this work
and conventional perturbation theory where the delta function
expressing energy conservation is replaced ad-hoc by a Lorentzian. 
Since the wave functions depend on $\Gamma$ only through an
exponential factor, the simple assumption that the total
decay-width is additive, is valid,
i.e. 
% sum of the decay-widths of the individual particles 
(as used e.g. in Ref. \cite{cyc}),
\begin{equation}
 \Gamma_{\rm tot} = \Gamma_{n_1} + \Gamma_{n_2} + \Gamma_r (s) \; .
\end{equation}
Hence, quantum correlations between different decaying states
exist only for non-Lorentzian spectral functions.

 %%%%%%%%%%%%

Second, let us consider magnetic Compton scattering as a 
typical second-order process. The electron propagator coincides 
with the conventional one with complexified
energy for  $Z_n (E) =Z^L_n (E)$. Therefore, in the 
simplest
case of $N_i =0 \to N_f =0$ Compton scattering where the energy of the 
photons is below $2m/\sin\theta$, we obtain the old, 
well-known result \cite{he79,bu86}. 
Otherwise, the instability of the initial
and final particles is also incorporated in the $S$-matrix element. 
Assuming the functions $Z^{1/2}$ are well-behaved, the integration over 
the off-shell energies will remove the remaining singularity of the electron
propagator when the virtual electron is in the stable Landau ground
state. In particular, the $S$-matrix element of magnetic Compton 
scattering is now finite in the case where the initial
photon is above the pair creation threshold.

Now we want to make our argument quantitative. The
$S$-matrix element of 
$N_i =0 \to N_f =0$ Compton scattering is given by
\begin{eqnarray}
 S_{fi}& = & (ie)^2 \int d^4 x  d^4 x'
          \bar\psi_f^{(+)}(x) \gamma_\mu iS_F (x,x') 
          \gamma_\nu \psi_i^{(+)}(x')
          \int ds^\prime_i ds^\prime_f  Z_{+,r_i}^{1/2} (s^\prime_i)  
                          Z_{-,r_f}^{1/2} (s^\prime_f)   
\nonumber\\ & &  
          \left(  A^{\mu\ast}_f (x,s^\prime_f)  A^{\nu}_i (x',s^\prime_i)
                + A^{\mu}_i (x,s^\prime_i)  A^{\nu\ast}_f (x',s^\prime_f) 
          \right)
         = S^{(1)}+S^{(2)}   \; ,           
\end{eqnarray}
where $i$ and $f$ refer to initial and final states, while the
the quantum numbers of the virtual electron will be marked by the
subscript $a$. Using stable fields,
the divergence for $s_i> 2m$ occurs in the exchange diagram $S^{(2)}$
when the virtual electron propagates as a positron in the Landau 
ground state,
i.e. has the quantum numbers $N_a=0$ and $\lambda_a =-1$ \cite{he79}. 
Since the space integrals remain unchanged and are finite, we
only have to consider the time integrals of $S^{(2)}$ for 
$N_a =0$ and $\lambda_a =0$,
\begin{eqnarray}
 S^{(2)} & \propto &  
       \int dt dt'
        e^{iE_f t} 
        \left(e^{iE_a (t-t')} \int\frac{dz}{2\pi}\: e^{iz(t-t')} \right)
         e^{-iE_i t'}
\nonumber\\ & &     
        \int ds^\prime_i ds^\prime_f Z_{+,r_i} ^{1/2} (s^\prime_i)  
                       Z_{-,r_f} ^{1/2} (s^\prime_f)   
         e^{-i \omega_i^\prime t'} e^{i\omega_f^\prime t}  
         =  S^{(2)}_t \; .
\end{eqnarray}
Here, $E$ denotes the energy of the electrons and $\omega'=s'\sin\theta$ 
the (off-shell) energy of the photons.
Performing the two time integrations and the integration over $z$, which
comes from the $\theta$-function of the electron propagator, results in
\begin{equation}
 S^{(2)}_t = \int ds^\prime_i ds^\prime_f 
           \:\frac{Z_{+,r_i} ^{1/2} (s^\prime_i) 
                   Z_{-,r_f} ^{1/2} (s^\prime_f)}
             {-E_f -E_a +\omega^\prime_i +i\varepsilon}\:
           2\pi \delta (E_f+\omega^\prime_f -E_i-\omega^\prime_i)   \; .   
\end{equation}
The result of the integration over $s^\prime_i$ depends on the energy 
perpendicular
to the magnetic field $s^\prime_f$ of the final photon,
\begin{eqnarray}
 S^{(2)}_t & = & 
         \frac{2\pi Z_{+,r_i}^{1/2} (s_{i,1}^\prime)} % Z_{-,r_f}^{1/2} (s_f)}
              {-E_i -E_a +\omega_f+i\varepsilon}\:  \sin\theta_f
          % 2\pi \delta (E_f+\omega_f -E_i- \omega^\prime_i)    
           \left| \begin{array}{c}
                   \\ s_f < 2m
                  \end{array} 
           \right. 
\nonumber\\ & &
         + \int  ds^\prime_f 
           \:\frac{2\pi Z_{+,r_i}^{1/2} (s_{i,2}^\prime) 
                        Z_{-,r_f}^{1/2} (s_f^\prime)}
                     {-E_i -E_a +\omega^\prime_f+i\varepsilon}\: 
           \sin\theta_i
           %\delta (E_f+s_f -E_i-s_i)     
           \left| \begin{array}{c}
                   \\ s^\prime_f \geq 2m
                  \end{array} 
           \right.  \; ,
\end{eqnarray}
where we introduce $s_{i,1}^\prime = (E_f+\omega_f-E_i)\sin\theta_i$,
$s_{i,2}^\prime = (E_f +\omega_f^\prime -E_i)\sin\theta_i$ and 
$\omega_f$ denotes
the on-shell energy of the final photon. 
In the second term of $S^{(2)}$,
the integration over $s^\prime_f$ gives a finite result as long as 
a principal value integral of the integrand 
%$Z_{+,r_i}^{1/2} (s_i^\prime) Z_{-,r_f}^{1/2} (s_f)$
can be defined.
Therefore, the only remaining dangerous part of $S^{(2)}$ is the first 
term. But for $s_f < 2m$ the denominator can never become zero because
$E_i + E_a=m+\sqrt{m^2+\omega_f^2\cos^2\theta_f}$ is always greater 
than $\omega_f$. Here, we assumed $p_{i,z}=0$ without loss of generality
and used momentum conservation parallel to the magnetic field.

In the more general case of Compton scattering when $N_i$ and $N_f$
are not restricted to be zero, this result remains valid. In this case,
there are additionally integrations over the off-shell energies of the
unstable electrons. As above, the necessary condition for a finite
$S$-matrix element is that a principal value integral of the integrand can 
be defined.

This example illustrates well the connection between the instability of
external particles and singularities of intermediate states:
as soon as the energy of the initial photon is above the pair creation
threshold, the virtual electron can become real, producing a divergent
cross section. At the same time, however, the photons also become
unstable. Taking this instability into account properly, 
one obtains well-behaved cross sections.

Finally, we want to comment on the behavior of magnetic Compton 
scattering if all external particles are stable. Then there is no 
decay-width which could cure the resonance if the virtual electron is in 
the Landau ground state $N_a =0$. Formally, the resonance energies 
$\omega_{\rm res}^{N_a}$ are given for all $N_a = 0,1,\ldots$
by \cite{he79}
\begin{equation}
 \omega_{\rm res}^{N_a} = 
      \left[ \left( m^2 +2N_a eB\sin^2\theta_i \right)^{1/2}
                                  -m \right] / \sin^2\theta_i  \; ,
\end{equation}
i.e. $\omega_{\rm res}^{N_a =0} =0$. In the case of $N_i \neq 0 \to N_f =0$
Compton scattering, the limit $\omega_i\to 0$ results in divergent cross
sections \cite{br}. In Ref. \cite{ka}, these divergences were
interpreted not as resonances but as infrared divergences.
By contrast, in the case of $N_i = 0 \to N_f =0$ scattering, the
$S$-matrix element diverges like $\omega_i^{-1}$ while the cross
section goes to a finite, constant value in the limit
$\omega_i \to 0$ \cite{br}.

%%%%%%%%%%%%%%%%%%%%%%%%%%%%%%%%%%%%%%%%%%%%%%%%%%%%%%%%%%%%%%%%%%%%%%%
\section{Summary}

We have presented a consistent method to describe the 
instability of excited Landau states and photons above the pair
creation threshold in QED of strong magnetic fields. This approach
consists in using Licht fields for unstable states and introduces 
additionally integrations over the off-shell energies of the
unstable particles.
We have shown for the generic example of Compton scattering
where the energy of the photon is above the pair creation threshold
how in this way the resonant divergences of $S$-matrix elements of
QED of strong magnetic field are avoided.

In the Breit-Wigner approximation the Licht states are exponentially
decaying or growing in time. Since the divergent part of the wave
functions is cut off
by $\theta$ functions, no divergent time integrals will be caused
by the real part of the exponentials. 
Therefore, in this formalism
%the instability of external states now can be taken into account and 
it is possible to calculate $S$-matrix
elements with external unstable particles.

%%%%%%%%%%%%%%%%%%%%%%%%%%%%%%%%%%%%%%%%%%%%%%%%%%%%%%%%%%%%%%%%
%\section{acknowledgments}
\acknowledgments

I am indebted to V. Berezinsky and G. Wunner
for continuous encouragement. 
This work was supported by grants from Deutscher Akademischer 
Austauschdienst and Land Nordrhein-Westfalen.

%%%%%%%%%%%%%%%%%%%%%%%%%%%%%%%%%%%%%%%%%%%%%%%%%%%%%%%%%%%%%%%%%


\begin{references}
\bibitem{spin} The propagator for a free Licht field with arbitrary spin 
               was given in Ref. \cite{lu}.
\end{references}
\end{document}